\documentclass[pra,letterpaper,twocolumn,superscriptaddress,floatfix]{revtex4}
\usepackage{graphicx,psfrag,amsmath,amssymb,amsfonts,bbm,latexsym,color,dcolumn,bm}

\begin{document}
	
\title{State dependence of tunneling processes and thermonuclear fusion}

\author{Roberto Onofrio}
\thanks{onofrior@gmail.com}

\affiliation{\mbox{Department of Physics and Astronomy, Dartmouth College, 6127 Wilder Laboratory, 
		Hanover, NH 03755, USA}}

\author{Carlo Presilla}
\thanks{carlo.presilla@uniroma1.it}

\affiliation{\mbox{Dipartimento di Matematica, Sapienza Universit\`a di
  Roma, Piazzale Aldo Moro 2, Roma 00185, Italy}}
  
\affiliation{Istituto Nazionale di Fisica Nucleare, Sezione di Roma 1, Roma 00185, Italy}

\begin{abstract}
  We discuss the sensitivity of tunneling processes to the initial
  preparation of the quantum state. We compare the case of Gaussian
  wave packets of different positional variances using a generalized
  Woods-Saxon potential for which analytical expressions of the
  tunneling coefficients are available. Using realistic parameters for
  barrier potentials we find that the usual plane wave approximation
  underestimates fusion reactivities by an order of magnitude in a
  range of temperatures of practical relevance for controlled energy
  production.
\end{abstract}

\maketitle

\section{Introduction}

Tunneling processes are of crucial relevance to a broad range of
physical systems, including semiconductors~\cite{Esaki} and
hererostructures~\cite{Vasko}, $\alpha$-radioactivity and nuclear
fusion in stars \cite{Gamow,Gurney,Adelberger}, the early
Universe~\cite{Atkatz}, and nuclear fusion processes in the laboratory
\cite{Bekerman1988,Vanderbosch1992,Balantekin1998,Hagino2022}.  Apart
from an early contribution~\cite{MacColl}, tunneling probabilities
have been usually evaluated by considering incoming plane waves.
However in realistic settings as the ones mentioned above, the
particles undergoing tunneling cannot in general be fully described by
plane waves, either because particles are confined in space, or
because in a many-body setting they undergo scattering with other
particles, thereby limiting the coherence length of the plane wave
\cite{Kadomtsev1997}. Moreover, there are discrepancies between
theoretical expectations and data from fusion experiments~\cite{Vaz},
and therefore it may be important to scrutinize all the underlying
theoretical assumptions.

It is therefore important to discuss the robustness of tunneling
coefficients and fusion reactivities with respect to the choice of
more general initial states, for instance by considering the
representative set of Gaussian wave packets.  The use of generalized
Gaussian wave packets has been already pioneered by Dodonov and
collaborators
\cite{Dodonov1996,Andreata2004,Dodonov2014a,Dodonov2014b}, with
results confirming that the predictions on tunneling rates may differ
even orders of magnitude with respect to the one arising from the
Wentzel-Kramer-Brillouin (WKB) approximation usually employed for
fusion reactivites. These studies, in particular~\cite{Dodonov2014a},
have been focused on analytical expressions valid under specific
conditions, not necessarily encompassing the entire parameter space.

The main goal of the present paper is to extend the above
  results evaluating the tunneling coefficient for arbitrary values of
  the position and momentum spreading.  The analysis is carried out
  having in mind applications to high-temperature ionized gases such
  as light nuclei plasmas in magnetically confined setups.  Fusion
  experiments with heavy-ion beams share the needs to incorporate the
  role of the energy width using a description in terms of wave
  packets \cite{Rubbia,Liu}.  A key ingredient of our discussion is
  the use of a potential admitting exact solutions for the tunneling
  coefficient in the entire energy range. This allows us to pinpoint
differences arising from the sole structure of the incoming Gaussian
wave packets, excluding other sources of differences as the ones due
to the use of approximations in the calculating techniques.
Additionally, we provide more intuitive arguments for the behavior of
fusion reactivity in both the cases of very narrow and very broad
positional variances.  We finally identify optimal
  operating temperatures for which reactivity gains with the
  corresponding wave packet states occur with respect to the case of
  plane wave states.

\section{Tunneling from wave packet states}

\begin{figure}[t]
  \begin{center}
    {\includegraphics[width=1.00\columnwidth,clip]{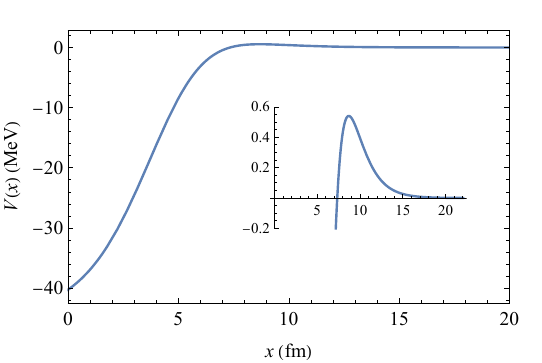}}
    \caption{Positive abscissa plot of the symmetric Generalized
      Woods-Saxon potential~\cite{GWS} experienced between two nuclei,
      with parameters $a=0.6~\mathrm{fm}^{-1}$, $L=5~\mathrm{fm}$,
      $V_0=45~\mathrm{MeV}$, and $W_0=56~\mathrm{MeV}$. With these
      parameters the barrier height (from zero to the maximum positive
      value of $V(x)$) is $0.540~\mathrm{MeV}$, the well width and
      depth are $7.35~\mathrm{fm}$ and $40.336~\mathrm{MeV}$,
      respectively.  All following figures are obtained using these
      parameters.  The inset (vertical units in MeV, horizontal units
      in fm) allows to better identify the shape of the barrier
      otherwise barely visible on the broader scale of the well
      depth.}
    \label{fig1}
  \end{center}
\end{figure}

We focus the attention on the Generalized Woods-Saxon (GWS) potential
energy for a one-dimensional system first introduced in~\cite{GWS}
(see also~\cite{Sever} for a simpler treatment)
\begin{eqnarray}
  V(x) = -\frac{V_0}{1+e^{a(|x|-L)}} +
  \frac{W_0 e^{a(|x|-L)}}{(1+e^{a(|x|-L)})^2} ,
\end{eqnarray}
where both $V_0$ and $W_0$ determine the peak values of the potential
energy, and $L, a$, as in the usual Woods-Saxon potential, determine,
respectively, the size of the effective well around the origin and its
spatial spread. For a convenient choice of these four parameters, the
GWS potential represents a symmetric well with value in the origin
equal to $-V_0/(1+\exp(-aL))+ W_0 \exp(-aL)/(1+\exp(-aL))^2$, and
$-V_0/2+W_0/4$ at $|x|=L$.  At large distances $|x| \gg L$ the
potential energy decreases exponentially to zero as
$V(x) \simeq (W_0-V_0) \exp(-ax)$, {\it i.e.}, within a range
$\lambda \simeq 1/a$.  This means that a semiqualitative difference
from potential energies of interest for instance in nuclear fusion is
that the barrier experienced by the nucleons, if schematized with this
potential, does not have the long range as expected for Coulomb
interactions, though in a realistic plasma the latter are screened on
the Debye length. In light of the simplicity of the potential
  capturing the essential feature of the tunneling process, and the
  presence of exact solutions, we do not expect the results being
  qualitatively different from the ones achievable by more
  sophisticated analyses.  We choose the set of parameters as
described in the caption of Fig.~\ref{fig1}, resulting in well depth,
barrier height and width of the well comparable to the ones of light
nuclei. We do not consider here the effect on tunneling of intrinsic 
degrees of freedom, such as vibrational or rotational couplings among the nucleons 
generating excited states, see \cite{Kimura}, therefore focusing only of the dependence 
of fusion rates on the external - due to translational motion - state. Using this potential and 
the related solutions in terms of tunneling coefficients $T(E)$ evaluated for plane waves at energy $E$, we have
considered more general cases of wave localized in both space and
momentum.  The most practical case, though not exhaustive of all
possibilities, is a Gaussian wave packet.

Let us consider the scattering of a one-dimensional Gaussian wave
  packet with positional variance $\xi^2$, characteristic wave vector
$K$ and mean energy $\hbar^2 K^2/(2m)$.  The corresponding
  wavefunction reads
\begin{equation}
  \psi(x)= \left(\frac{2}{\pi \xi^2}\right)^{1/4} e^{-(x-x_0)^2/\xi^2+i K x},
\end{equation}
which in wave vector space $k$ becomes
\begin{eqnarray}
  \varphi(k)
  &=&
      \frac{1}{\sqrt{2\pi}}
      \int_{-\infty}^{+\infty} \psi(x) e^{-i k x} dx
      \nonumber \\
  &=&
      \frac{1}{(2 \pi)^{1/4}} \sqrt{\xi}
      e^{-\xi^2 (k-K)^2/4} e^{i(K-k)x}.
\end{eqnarray}
This equation shows that we are dealing with an ensemble of plane
  waves with wave vector $k\in (-\infty,+\infty)$ distributed
  according to the probability density
\begin{equation}
  \label{PkK}
  P(k,K) = \vert \varphi(k)\vert^2
  = \frac{\xi}{\sqrt{2 \pi}} e^{-\xi^2 (k-K)^2/2},
\end{equation}
where we have introduced the positional spreading $\xi$ as the square
root of the positional variance.  Note that the probability
  density $P(k,K)$ is a Gaussian function of $k-K$, i.e., it depends
  on the mean energy $\hbar^2 K^2/(2m)$ of the packet via its
  characteristic wave vector $K$. 
  
   \begin{figure}[t]
    \begin{center}
      {\includegraphics[width=1.00\columnwidth,clip]{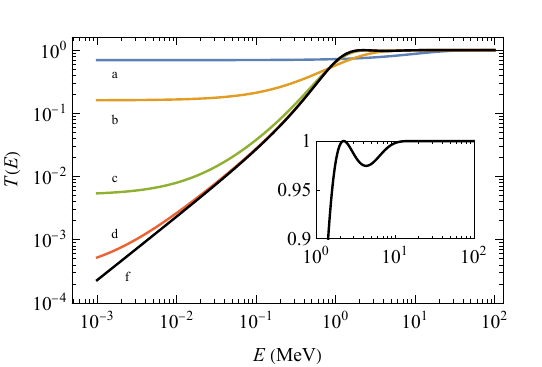}}
      \caption{Transmission coefficient of a Gaussian wave packet of
        width $\xi$ and mean energy $E$ impinging on the GWS
        potential of Fig.~\ref{fig1}. Curves from a to d respectively
        correspond to the cases $\xi=2 ,8, 32,128~\mathrm{fm}$, while
        the case of tunneling of a plane wave f is also depicted.
        In the inset we show a zoom-in of curve f on the region
          close to complete transmission, clearly showing one resonant
          tunneling oscillation. Note that this oscillation is
          progressively smeared out in curves d-a by decreasing
          $\xi$.}
      \label{fig2}
    \end{center}
  \end{figure}
  
\section{Fusion reactivities from thermal states}
 
The characteristic wave vector $K$ around which the wave vectors of the packet are 
distributed according to quantum mechanical probabilities is in turn 
distributed according to the initial classical preparation of the particle's energies.
Equation~(\ref{PkK}) allows us to consider fusion processes taking place 
in an ensemble of nuclei represented by Gaussian wave packets with 
arbitrary distribution of their mean energy. In particular, apart from the 
case of a monoenergetic distribution, or a highly peaked one, therefore represented by 
a single or a narrow range of $K$ values, respectively -- as typically realized 
in fusion experiments with ion beams -- it is important to consider a Maxwell-Boltzmann 
energy distribution of the particles when modelizing thermonuclear fusion. 
This translates into considering the convolution of the $k$ wave vectors quantum mechanical distribution
and the $K$ characteristic wave vector classical distribution, reminiscent of the 
Voigt profile broadly used in atomic spectroscopy \cite{Voigt1,Voigt2}.
Therefore we now assume that the nuclei are at thermal canonical
  equilibrium with inverse temperature $\beta$. The one-dimensional
  wave vector $K$ is then distributed according to the Maxwell-Boltzmann
  probability density
\begin{equation}
  \label{wMB}
  w(K,\beta)_{MB} = \left(\frac{\beta}{\pi} \frac{\hbar^2}{2m} \right)^{1/2}
  e^{-\beta \hbar^2 K^2/(2m)},
\end{equation}
where $m=m_a m_b/(m_a+m_b)$ is the reduced mass of the two 
nuclei $a$ and $b$ which actually take part in the fusion process.
Eq.~(\ref{wMB}), in which the wave vector $K$ can assume any 
positive or negative value, is normalized as
  \begin{equation*}
    1 = \int_{-\infty}^{+\infty} w(K,\beta)_{MB} ~dK.
  \end{equation*}
The spread of the wave vector $K$ is determined by the inverse temperature 
$\beta$, as customary for canonical ensembles.

  \begin{figure}[t]
    \begin{center}
      {\includegraphics[width=1.00\columnwidth,clip]{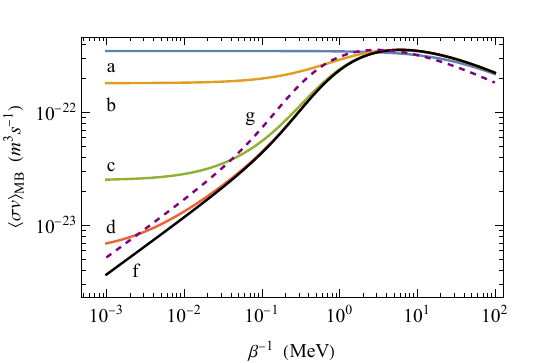}}
      \caption{Fusion reactivity, evaluated for the same Gaussian wave
        packets of Fig.~\ref{fig2} (and same labeling a-f) with energy
        $E$ averaged over a Maxwell-Boltzmann distribution as a
        function of the temperature $\beta^{-1}$. The dashed line
        (case g) is the case in which the positional spreading depends
        on temperature as $\xi=\lambda(\beta)/\sqrt{2}$, where
        $\lambda(\beta)$ is the thermal wavelength of the
        nuclei~\cite{Chenu}. The ratio between this latter curve and
        the curve for a plane wave (case f) versus $\beta^{-1}$ is
        reported in the inset to evidence their differences in a
        region of interest for nuclear fusion.}
      \label{fig3}
    \end{center}
  \end{figure}
  
  Particularly relevant to the discussion of fusion
    processes is the reactivity defined as
  $\langle \sigma(E) v(E) \rangle_{MB}$ where
  $\langle ... \rangle_{MB}$ denotes the average over the
    statistical distribution of the reactants, which in our case is
    the Maxwell-Boltzmann distribution, $\sigma(E)$ is the cross-section
    of the process, and $v(E)$ is the particle velocity.
    In one dimension and for nuclei
    at energy $E=m v^2/2=\hbar^2 k^2/(2m)$, the cross-section is
    $\sigma(E) = T(E) \pi / k^2$ \cite{Satchler}, and 
  \begin{equation}
    \sigma(E) v(E) =
    \frac{\pi \hbar^2}{\sqrt{2 m^3}} \frac{1}{\sqrt{E}} T(E).
  \end{equation}

 \begin{figure}[t]
    \begin{center}
      {\includegraphics[width=1.00\columnwidth,clip]{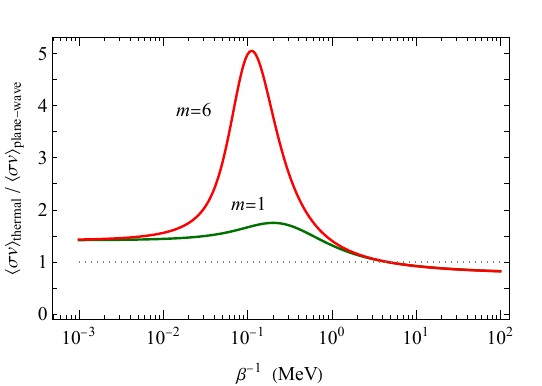}}
      \caption{Enhancement of fusion reactivity with 
        thermal wave packets. The ratio between the dashed line 
        (case g in Fig.~\ref{fig3}) in which the positional spreading depends
        on temperature as $\xi=\lambda(\beta)/\sqrt{2}$, where
        $\lambda(\beta)$ is the thermal wavelength of the
        nuclei~\cite{Chenu}, and the curve for a plane wave (case f in  Fig.~\ref{fig3}), 
        is reported versus $\beta^{-1}$ in a temperature region of interest for nuclear fusion. 
        Two cases allow to evidence the dependence of the enhancement on the reduced mass of the nuclei.}
      \label{fig4}
    \end{center}
  \end{figure}  
  In our case, each nucleus does not correspond to a monochromatic
    plane wave at energy $E=\hbar^2k^2/(2m)$, instead it is
    represented by a Gaussian wave packet with wave vectors $k$
    distributed with probability density $P(k,K)$.  In turn, the
    Gaussian wave packets representing the nuclei in the thermal
      ensemble have characteristic wave vectors $K$ spread with
    Maxwell-Boltzmann distribution $w(K,\beta)_{MB}$. It follows that
    the average fusion reactivity is
  \begin{eqnarray}
    \langle \sigma v \rangle_{MB}
    &=&
        \frac{\pi \hbar^2}{\sqrt{2 m^3}}
        \int_{-\infty}^{+\infty} dk \int_{-\infty}^{+\infty} dK
        \left(\frac{\hbar^2 k^2}{2m}\right)^{-1/2}
        \nonumber\\
    && \quad\times~
       T(\frac{\hbar^2 k^2}{2m}) P(k,K) w(K,\beta)_{MB}.
       \label{reactivity2}
  \end{eqnarray}
  Due to the Gaussian nature of the functions $P(k,K)$ and
    $w(K,\beta)_{MB}$, the integral over $K$ can be evaluated
    analytically, yielding the rather compact formula
  \begin{equation}
    \langle \sigma v \rangle_{MB}= 
    \frac{\sqrt{\pi}}{2} \frac{\hbar}{m}
    \int_{-\infty}^{+\infty} dk \frac{1}{k} T({\frac{\hbar^2 k^2}{2m}})
    \xi_{\mathrm{eff}} e^{-\xi_{\mathrm{eff}}^2k^2/2},
    \label{reactivity3}
  \end{equation}
  where we have introduced an effective positional spreading
  $\xi_{\mathrm{eff}}$, depending on the inverse temperature, such
  that
  \begin{equation}
    \frac{1}{\xi_{\mathrm{eff}}^2} = \frac{1}{\xi^2}+\frac{m}{\beta \hbar^2}.
    \label{xieff}
  \end{equation}
  States approximating a plane wave satisfy
    $\xi^2 \gg \beta \hbar^2/m$, therefore
    $\xi_{\mathrm{eff}}^2 \simeq \beta \hbar^2/m$, i.e.,
    $\xi_{\mathrm{eff}}$ becomes the thermal De Broglie wavelength.
    In the opposite limit of states highly localized in position,
    $\xi^2 \ll \beta \hbar^2/m$, we have
    $\xi_{\mathrm{eff}} \simeq \xi$.  High-temperature Boltzmann
    states are then approximating wave vector eigenstates (of
    eigenvalue $K$), while low-temperature Boltzmann states
    approximate position eigenstates.  This shows that even assuming
  an initial quantum state with positional variance of quantum nature,
  at temperature large enough the relevant lengthscale below which
  quantum coherence of the wave packet is maintained no longer depends
  on the initial preparation. Analogous conclusions have been already
  obtained in~\cite{Alterman,Chenu}. This can also be interpreted, in
  the case of a gas at given temperature and density, as corresponding
  to the mean free path for two-particle collisions. 
  
     \begin{figure}[t]
    \begin{center}
      {\includegraphics[width=1.00\columnwidth,clip]{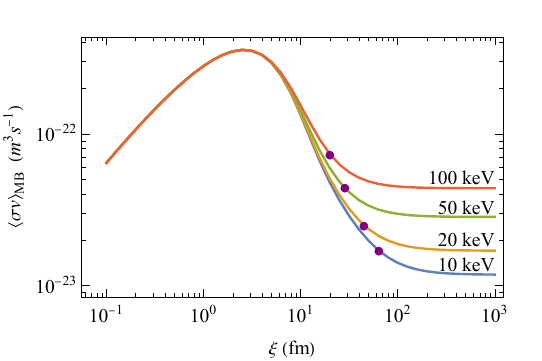}}
      \caption{Fusion reactivity versus the positional spreading $\xi$
        for four temperature values $\beta^{-1}=10,20,50,100$ keV of
        relevance in fusion of light nuclei. The dots denote the
        reactivities for the positional spreading from the thermal
        wavelength of the nuclei as discussed in Fig.~\ref{fig3} and
        evaluated at the corresponding temperatures shown here.}
      \label{fig5}
    \end{center}
  \end{figure}
  
  The tunneling coefficient versus the mean energy of the wave
  packet $E$ is depicted in Fig.~\ref{fig2} for various values of the
  width $\xi$ of the Gaussian wave packet. The dependence of the
  tunneling coefficient on $E$ is, quite predictably, mild when the
  value of $E$ is comparable or higher than the barrier
  height. Instead its dependence at lower energies strongly depends on
  $\xi$, with the case of plane waves (in the limit of
  $\xi \rightarrow +\infty$) underestimating the transmission
  coefficient by even five orders of magnitude at the lowest reported
  energies, with respect to the case of a Gaussian wave packet with
  size $\xi$ smaller than the size of the effective well. The case of
  small positional variance should correspond, for a state of minimal
  quantum uncertainty, to a broad distribution of possible momenta,
  including some corresponding to kinetic energies comparable or
  higher than the barrier height.  Notice the presence of resonant
  tunneling in the case of plane waves and spatially delocalized
  Gaussian wave packets, which is instead washed out in the
  integration when considering Gaussian wave packets of smaller width
  in position, and therefore broader in momentum/wave vector space.

  In Fig.~\ref{fig3} we present the average reactivity
  corresponding to a Maxwell-Boltzmann distribution versus temperature
  for different values of the positional spreading $\xi$. Reflecting
  the results presented in Fig.~\ref{fig2}, the high temperature
  behavior is the same for the various cases, while at low temperature
  the same pattern appears, with the highest reactivity occurring for
  the Gaussian wave packet of smallest value. Notice a further curve
  (dashed) which is evaluated for a temperature-dependent positional
  spreading as discussed in~\cite{Chenu}.  This curve is relevant for
  at least two reasons.  First, without any active control of the
  positional variance of the wave packet, this is what we expect by
  considering a gas of reagents with a Maxwell-Boltzmann
  distribution. Secondly, in the temperature range between 10 keV and
  100 keV, of interest for controlled thermonuclear fusion, we
  estimate a boost of the reactivities if compared to the ones
  achieved by considering plane waves.    
 This is more easily noticeable in Fig.~\ref{fig4}, where we report the  
  ratio between the dashed curve of Fig.~\ref{fig3} and the curve corresponding to the prediction of
  plane waves, versus the temperature. In the above
  mentioned range the ratio is about 1.5, followed by a mild increase
  to almost 2, then becoming smaller than unity at even higher
  temperatures. The peak value of the ratio depends on the involved
  masses, as shown in the comparison of the two nucleons with a mass
  of 2 a.m.u. (reduced mass of 1 a.m.u.) and 12 a.m.u. (reduced mass
  of 6 a.m.u.). While the latter example has been chosen having in
  mind the case of Carbon quite relevant in astrophysics, it should be
  kept in mind that the same GWS potential is used in both cases to
  see the sole dependence on the mass, which is unrealistic for Carbon
  especially in regard to its actual larger well width.

We emphasize more these considerations from a complementary standpoint
by plotting the reactivity as a function of the positional variance
$\xi$ for values of temperature relevant to fusion processes of light
nuclei, $\beta^{-1}=10, 20, 50, 100$ keV, as depicted in
Fig.~\ref{fig5}.  This plot allows to better appreciate that there is
an optimal value of $\xi$ maximizing the reactivity at a given
temperature.  Indeed, in the case of $\xi \rightarrow 0$ there will be
increasing components of the wave packet at large $k$. These
components will saturate the transmission coefficient to its maximum
value, and will strongly suppress the cross-section due to the
dependence of the latter upon $1/k^2$, with the overall dependence on
reactivity then scaling as the inverse of the wave vector.

The above results have been tested for various choices of the
parameters of the potential with outcome qualitatively similar to the
specific case considered in this paper.  We expect robustness also in
the case of a potential which is the sum of a flat potential at
distances smaller than the average radius of the nuclei, and a Coulomb
potential. The outcome should also hold in the more realistic
three-dimensional setting, when including effects due to the angular
momentum, and a spherically symmetric electric field inside the
nucleus assuming uniform electric charge density.  
  Indeed, intuitively, in the full three-dimensional case the
  scattering of plane waves as initial states is expected to differ
  even more from the case of initial states well localized in
  position, and the thermal distribution of $K$ will be proportional
  to $K^2 \exp(-\beta \hbar^2 K^2/(2m))$. However, more extensive
analyses will be necessary to determine the quantitative gain in using
optimized Gaussian wave packets under these more realistic -- yet not
susceptible of analytic solutions -- situations.

\section{Conclusions}

In conclusion, we have investigated the sensitivity of tunneling
processes to the preparation of Gaussian wave packets -- and
contrasted to the usually assumed case of plane waves -- in the case
of an analytically solvable potential  and in the presence of 
a canonical ensemble of nuclei. Two sources of uncertainty in the 
knowledge of the wave vector are present, the quantum mechanical uncertainty due to 
the consideration of a wave packet of positional spread $\xi$ instead of a plane wave, 
and the classical uncertainty in the energy of the particle belonging to an ensemble 
with an energy distribution given by classical statistical mechanics. In the specific case 
of Gaussian wave packets and a canonical distribution, the two sources of uncertainty 
are combined in a simple formula leading to a positional spreading as in Eq. (\ref{xieff}), 
which can also be interpreted as a positional spreading renormalized by the presence of  
the environment of surrounding nuclei at finite temperature. 
We have evidenced sensitivity of the resulting reactivities for fusion processes, a
  result of interest also in the astrophysical setting for primordial
  nucleosynthesis \cite{Bonetti}. It is still unclear how to engineer in general 
  wave packets of well-defined, targeted, positional variance.  
 These results should provide further stimuli to design thermonuclear fusion 
 prototypes in which emphasis is put in maximizing the plasma temperature 
 with more moderate plasma density, an important point for achieving 
 deuterium-deuterium fusion, with well-known advantages with respect to 
 the currently experimentally investigated deuterium-tritium fusion \cite{Onofrio}.

\end{document}